\newcommand{\beq}{\begin{equation}}
\newcommand{\eeq}{\end{equation}}
\newcommand{\bea}{\begin{eqnarray}}
\newcommand{\eea}{\end{eqnarray}}
\newcommand{\nn}{\nonumber}

\newcommand{\rh}{\rho}

\newcommand{\bv}{{\mathbf v}}
\newcommand{\bB}{{\mathbf B}}
\newcommand{\dadb}[2]{\frac{{  d}#1}{{  d}#2}}

\documentclass[twocolumn,amsmath,amssymb,aps]{revtex4-1}

\usepackage[utf8]{inputenc}
\usepackage[T1]{fontenc}
\usepackage{amsmath}

\usepackage{mathtools}
\usepackage{amsfonts}
\usepackage{graphicx}
\usepackage{dcolumn}
\usepackage{bm}
\usepackage[export]{adjustbox}
\usepackage{wrapfig}
\usepackage{lipsum}
\usepackage[titletoc,toc,title]{appendix}
\usepackage[colorlinks]{hyperref}
\usepackage{color}

\begin{document}

\title{ A tokamak pertinent analytic equilibrium  with plasma flow of arbitrary direction}

\author{D. A. Kaltsas$^1$}
 \email{dkaltsas@cc.uoi.gr}

 \author{A. Kuiroukidis$^2$}
 \email{apostk@teiser.gr}

\author{G. N. Throumoulopoulos$^1$}%
 \email{gthroum@uoi.gr}
\vspace{3mm}

\affiliation{\it   $^1$ Department of Physics, University of Ioannina,
   GR 451 10 Ioannina, Greece\\
   $^2$ Department of Informatics,
Technological Education Institute of Serres, GR 62124 Serres, Greece
  }


\begin{abstract}

An analytic solution to a generalized Grad-Shafranov equation with flow of arbitrary direction is obtained upon adopting the generic linearizing ansatz for the free functions related to the poloidal current, the static pressure and the electric field. Subsequently, a D-shaped tokamak pertinent equilibrium with sheared flow is constructed using the aforementioned solution.

\end{abstract}

\maketitle


The axisymmetric magnetohydrodynamic equilibrium states are governed by the widely known Grad-Shafranov (GS) equation. Analytic solutions to the generic linearized form of this equation was obtained in terms of infinite series involving confluent hypergeometric functions \cite{pfre,atgu}. These are solutions to the ordinary GS equation i.e. they can describe static equilibrium states or equilibria with incompressible flows parallel to the magnetic field. However, it is well known that macroscopic plasma flows, being either externally driven or intrinsic, are present in tokamaks (e.g. see \cite{Rice2016}). Plasma rotation affects the stability properties of the equilibria and moreover sheared flows contribute in the reduction of turbulent transport thereby playing a role in the transition to the advanced confinement regimes (L-H transitions). Therefore, generalized Grad-Shafranov equations  (GGSE), taking into account macroscopic flows, were also derived, e.g. Eq. \eqref{ggs} below. Aim of the present research note is to construct an analytic  solution to the generic linearized form of Eq. \eqref{ggs} and 
in addition to exploit this solution in constructing tokamak relevant stationary equilibria with an imposed boundary.

The MHD equilibrium states of an axisymmetric plasma with incompressible flow are governed by the following Alfv\'en normalized GGSE
   \cite{T-Th,simi,thta}
\begin{eqnarray}
\label{ggs}
&\Delta^{*}u+\frac{1}{2}\frac{d}{du}\left[\frac{X^{2}}{1-M_{p}^{2}}\right]
+\rho^{2}\frac{dP_{s}}{du}\nn\\
&+ \frac{\rho^{4}}{2}\frac{d}{du}
\left[\mu\left(\frac{d\Phi}{du}\right)^{2}\right]=0
\end{eqnarray}
Here, the function $u(\rho,\xi)$  labels the magnetic surfaces and  $\rho=R/R_{0}$, $\xi=z/R_{0}$ are normalized cylindrical coordinates ($R,\phi, z$) with $z$ corresponding to the axis of symmetry;  $\Phi(u)$ is
 the electrostatic potential and $\mu(u)$ the plasma density;
 $M_p(u)$ is
 the   Mach function of the poloidal fluid velocity with respect to the
poloidal  Alfv\'en velocity;
 $X(u)$ relates to the toroidal magnetic
 field, $B_\phi=I/R$,  through 
 \begin{eqnarray}
 I=\frac{X}{1-M_p^{2}}-\frac{\rho^2\sqrt{\mu}M_p d\Phi/du}{\sqrt{1-M_p^2}}
 \end{eqnarray} 
For vanishing flow the surface function $P_s(u)$
  coincides with the pressure; $B$ is the magnetic field modulus;
and $\Delta^\star=\rho^2\nabla\cdot(\nabla/\rho^2)$. Also, the velocity is decomposed to a component parallel to $\bB$ and a non-parallel one associated with the electric field:
\beq
\bv=\frac{M_p}{\sqrt{\mu}}\bB-\rho^2\left(1-M_p^2\right)^{1/2}\left(\dadb{\Phi}{u}\right)\nabla\phi
\label{vel}
\eeq

%

Adopting the linearizing ansatz for the free function terms,
\bea
\label{ansatz}
\frac{X^{2}}{2(1-M_{p}^{2})}&=&\frac{X_{0}^{2}}{2}+X_{1}u+\frac{1}{2}X_{2}u^{2}\nn\\
P_{s}&=&P_{0}+P_{1}u+\frac{1}{2}P_{2}u^{2}\nn \\
 \frac{1}{2} \left[\mu\left(\frac{d\Phi}{du}\right)^{2}\right]&=&G_{0}+G_{1}u+ \frac{1}{2}G_2u^{2}
\eea
results in the most general linear form of \eqref{ggs}:
\bea
\label{ggs2}
&u_{\rho\rho}-(1/\rho)u_{\rho}+u_{\xi\xi}+X_{1}+X_{2}u\nn\\
&+\rho^{2}(P_{1}+P_{2}u)+\rho^{4}(G_{1}+G_{2}u)=0
\eea
 Also,  ansatz \eqref{ansatz} permits two out of the five free functions involved to remain arbitrary. 
We will construct a solution to \eqref{ggs2} of the form
\bea
\label{ansatz1}
u:=u_{inh}(\rho)+U(\rho,\xi)
\eea
therefore Eq. \eqref{ggs2} is satisfied when the following equations for $u_{inh}(\rho)$ and $U(\rho, \xi)$ do so:
\bea
\label{geq1}
&u_{inh}^{''}-(1/\rho)u_{inh}^{'}+X_{1}+X_{2}u_{inh}+\rho^{2}P_{1}+
\rho^{2}P_{2}u_{inh}\nn\\
&+\rho^{4}G_{1}+\rho^{4}G_{2}u_{inh}=0
\eea
\bea
\label{geq2}
U_{\rho\rho}-(1/\rho)U_{\rho}+U_{\xi\xi}+X_{2}U+\rho^{2}P_{2}U+\rho^{4}G_{2}U=0
\eea

First, to solve Eq. \eqref{geq2} as in Refs. \cite{srin,Cerfon2010} we adopt  an expansion of the form
\bea
\label{expans1}
U(\rho,\xi)=\sum_{n=0}^{\infty}f_{n}(\rho)\xi^{n}
\eea
Substituting this into Eq. (\ref{geq2}) leads after multiplying by $\rh^2$ to
\bea
\label{general1}
\rho^{2}f_n^{''}-\rho f_n^{'}+(X_{2}\rho^{2}+P_{2}\rho^{4}
+G_{2}\rho^{6})f_n=\nn\\-
(n+1)(n+2)\rho^{2}f_{n+2}\,,\quad n=0,1,2,...
\eea

The general solution of this  equation is the sum of the general solution
to the homogeneous equation
\bea
\label{homog}
\rho^{2}y^{''}-\rho y^{'}+(X_{2}\rho^{2}+P_{2}\rho^{4}+G_{2}\rho^{6})y=0
\eea
(with $y:=f_n$)  plus a particular solution to the 
inhomogeneous equation.
To solve  \eqref{homog}
we will employ the Frobenius method, according to which its solution can be expressed  in the form of a convergent series around the regular singular point $\rho=0$:
\bea
\label{form1}
y=\sum_{n=0}^{\infty}a_{n}\rho^{n+r}
\eea
The index equation is $r(r-2)=0$ so that $r_{1}=2,\; r_{2}=0$. For $r_1=2$, substituting $y_1=\sum_{n=0}^{\infty}a_n\rho^{n+2}$ into \eqref{homog} yields the following recurrence relations for the coefficients $a_n$:
\bea
\label{rec1}
a_{2n+1}=0\,, \quad n&=&0,1,2,... \nn\\
 8a_{2}+X_{2}a_{0}&=&0 \nn\\
 24a_{4}+ X_{2}a_{2}+P_{2}a_{0}&=&0  \nn\\
n(n+2)a_{n}+  X_{2}a_{n-2}+P_{2}a_{n-4}&&\\
+G_{2}a_{n-6}&=&0\,, \quad n=6,8,10,...\nn
\eea
where $a_0\neq 0$ is arbitrary. Note that \eqref{rec1} imply that all the odd terms in \eqref{form1} vanish.
Since $r_{1}-r_{2}=2\in Z$  the second  independent solution
$y_{2}$ should be  of the form
\bea
\label{secsol}
y_{2}=ky_{1}log(\rho)+\sum_{n=0}^{\infty}A_{n}\rho^{n}
\eea
Equation \eqref{homog} then leads to the following recurrence relations for the coefficients $A_n$:
\bea
\label{rec2}
A_{2n+1}&=&0\, \quad n=0,1,2,...\nn \\
X_{2}A_{0}+2ka_{0}&=&0\nn \\
8A_{4}+X_{2}A_{2}+P_{2}A_{0}+6ka_{2}&=&0\nn \\
n(n+2)A_{n+2}+X_{2}A_{n}+P_{2}A_{n-2}&&\\+G_{2}A_{n-4}+2k(n+1)a_{n}&=&0\,,\quad
n=4,6,8,...\nn
\eea
where $A_2$ and $k$ remain arbitrary. Therefore, the general solution of \eqref{homog} is written as
\beq  y=C y_1 +D y_2 \label{generic} \eeq.

Turning now to a particular solution of \eqref{general1} we assume that
$f_{n}(\rho):=0$ for $ n \geq  N$,
implying that \eqref{general1} becomes homogeneous for $n =N-1$ and  $n= N-2$. Therefore, the inhomogeneous ``source'' term in \eqref{general1} for $n=N-3$  is known in terms of the generic solution of \eqref{homog}. Also this requirement makes \eqref{expans1} as a sum of $N$ terms instead of infinite to be exact. In view of this generic solution we pursue a particular solution of \eqref{general1} of the form
\begin{eqnarray}
y^p_{N-3}=AY_1 +B Y_2
\end{eqnarray}
where 
\bea
&Y_1=\sum_{m=0}^{\infty} b_m\rho^{m+2} \nn\\
&  Y_2=\log \rho \sum_{m=0}^\infty c_m \rho^{m+2}+\sum_{m=0}^\infty d_m \rho^{m}
\eea
 Demanding that $Y_1$  and $Y_2$ are solutions to \eqref{general1}  for $n=N-3$  we obtain the recurrence relations for the coefficients $b_m$:
\begin{eqnarray}
 b_{2m+1}&=&0\,, \quad m=0,1,2... \nn\\
8b_2+X_2 b_0+a_0&=&0\nn\\
24 b_4+X_2b_2+P_2b_0+a_2&=&0\nn\\
m(m+2)b_m+X_2b_{m-2}+P_2b_{m-4}&&\\
+G_2b_{m-6}+a_{m-2}&=&0\,, \quad m=6,8,10,...\nn
\end{eqnarray}
and similar relations for $c_m$ and $d_m$.
 Then, the general solution of \eqref{general1} (for $n=N-3$)   is
\begin{eqnarray}
y_{N-3} = Cy_1+Dy_2 +y^p_{N-3}
\end{eqnarray}
 By a similar procedure can  be found particular solutions of the inhomogeneous equations \eqref{general1} for $n=N-4,\ N-5,\ldots ,0$. 

\begin{figure}[h]
\begin{center}
\includegraphics[scale=0.6]{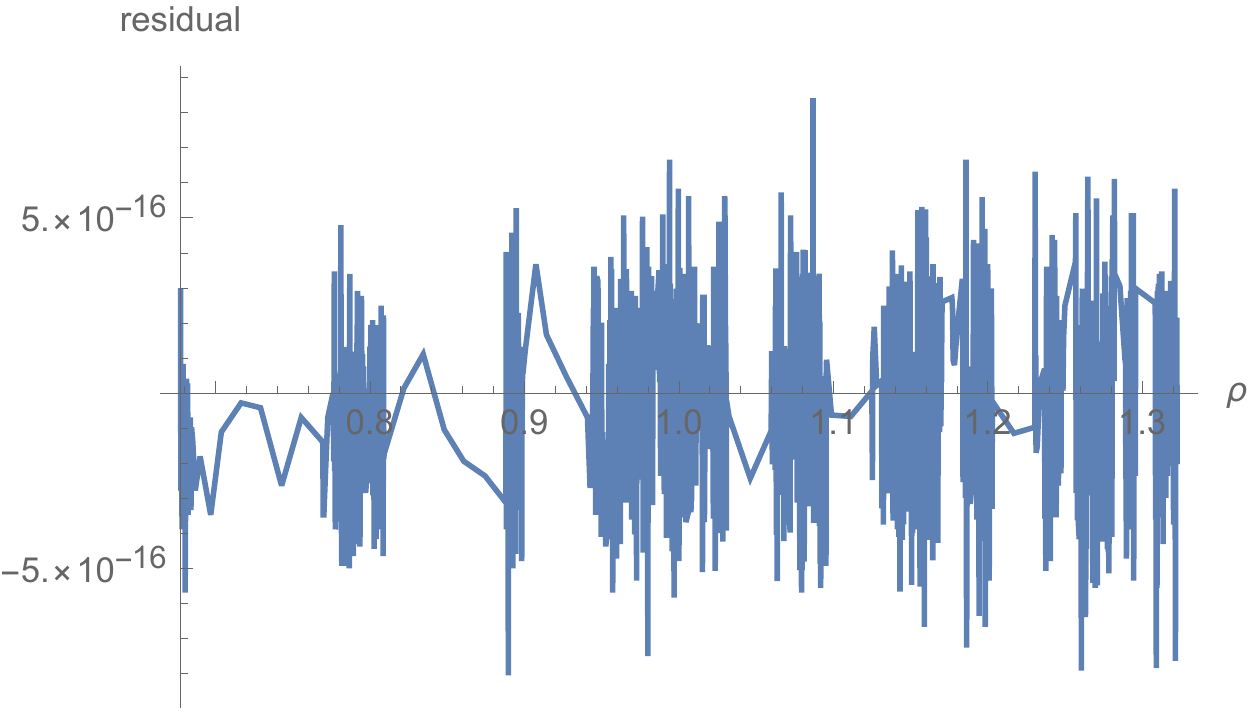}
\end{center}
\caption{The residual error of the power series solution for the Tokamak equilibrium of Fig. \ref{fig_2} on the equatorial plane $z=0$, indicates that \eqref{ggs2} is satisfied up to machine precision. \label{fig_1}}
\end{figure}

Finally, to solve Eq. \eqref{geq1} we consider a particular solution of the form 
\beq 
u_{inh}=\sum_{n=0}^\infty\, a_n(\rh-1)^n   \label{part2} 
\eeq 
Although the homogeneous solution was found upon expanding around the regular singular point $\rho=0$ (the Frobenius method guarantees that the convergence radius goes to infinity), we expand the inhomogeneous, particular solution around the regular point $\rho=1$. It turned out that this choice minimizes the residual error, i.e., the output obtained upon inserting the truncated solution into the lhs of Eq. \eqref{ggs2}, more efficiently. As a matter of fact we achieved a residual error of the order of the machine epsilon for $M\geq 30$ (see Fig. \ref{fig_1}), where $M$ is the number of terms in the various truncated series. This power series solution has radius of convergence at least up to the first singular point, i.e. $\rho=0$. Therefore, expansion \eqref{part2} is appropriate for our purpose i.e. the construction of non-compact tokamak equilibria.
The coefficients $a_n$ can be obtained in a similar recursive manner after demanding that \eqref{part2} satisfies \eqref{geq1} although the expansion around $\rho=1$ makes the analysis somewhat more involved.  \\ \indent
The above described algorithms were  implemented in developing  a Mathematica code solving Eq. \eqref{ggs}. Thus, we also checked the convergence of the  infinite series representing the particular solutions. As an example,   we construct  below D-shaped tokamak pertinent equilibria. We chose $N=6$ and included 33 terms in all the series. In order to construct equilibrium configurations with D-shaped boundaries we need to compute the arbitrary parameters of our final solution by imposing appropriate boundary conditions at a series of boundary points. Crucial for the success of this methodology is the number of points to be exploited, that is, usually the boundary shape is decently reproduced if this number is sufficiently large. However, upon employing a shaping method introduced in \cite{Cerfon2010} and utilized also in  \cite{Throumoulopoulos2012,Kaltsas2014} we can reduce the number of points that have to be used, in principle to three for up-down symmetric configurations and even solutions in $\xi$ and to four for asymmetric configurations or/and non-even solutions in $\xi$. This set of shaping conditions incorporates equations concerning the boundary values of the flux function $u(\rho,\xi)$ and its first and second order derivatives at  the top,  lower,  inner and  outer points of the configuration given by the coordinate pairs $(\rho_t,\xi_t)=(1-\delta\epsilon,\kappa\epsilon)$, $(\rho_d,\xi_d)=(1-\delta\epsilon,-\kappa\epsilon)$, $(\rho_i,\xi_i)=(1-\epsilon,0)$, $(\rho_o,\xi_o)=(1+\epsilon,0)$ respectively, where $\epsilon=a/R_0$ is the inverse aspect ratio of the tokamak, $\delta$, is the triangularity and $\kappa$ is the elongation along the $z$ axis. The justification for the utilization of these conditions is given in  \cite{Cerfon2010} and \cite{Throumoulopoulos2012, Kaltsas2014};  therefore we omit here  further discussion regarding their origin. The shaping conditions are summarized as follows
\begin{eqnarray}
&&u(1-\epsilon,0)=u(1+\epsilon,0)\nn\\
&&=u(1-\delta\epsilon,\kappa\epsilon)=u(1-\delta\epsilon,-\kappa\epsilon)=0\, \\
&&u_\rho(1-\delta\epsilon,\kappa\epsilon)=u_\rho(1-\delta\epsilon,-\kappa\epsilon)=0\,\\
&&u_\xi(1-\epsilon,0)=u_\xi(1+\epsilon,0)=0\\
&& u_{\xi\xi}(1-\epsilon,0)+
\frac{(1-\alpha)^2}{\epsilon\kappa^2}u_{\rho}(1-\epsilon,0)=0\,\\
&& u_{\xi\xi}(1+\epsilon,0)
-\frac{(1+\alpha)^2}{\epsilon\kappa^2}u_{\rho}(1+\epsilon,0)=0\, \\
&& u_{\rho\rho}(1-\delta\epsilon,\kappa\epsilon)-\frac{\kappa}{\epsilon\cos^2\alpha}u_{\xi}(1-\delta\epsilon,\kappa\epsilon)=0\,\\
&& u_{\rho\rho}(1-\delta\epsilon,-\kappa\varepsilon)-\frac{\kappa}{\epsilon\cos^2\alpha}u_{\xi}(1-\delta\epsilon,-\kappa\epsilon)=0\,
 \label{shaping}
\end{eqnarray}
This set of conditions,  being  sufficient to determine a maximum of twelve shaping parameters in our final solution,  is able to produce an up-down asymmetric D-shaped equilibrium configuration if $\kappa$ and $\delta$ are chosen differently for the upper and the lower parts or if the last shaping condition is replaced by $u_{\xi\xi}(1-\delta\epsilon,-\kappa\epsilon)=0$ for a diverted boundary with lower X-point.
\begin{figure}[h]
\begin{center}
\includegraphics[height=9cm,width=5.5cm]{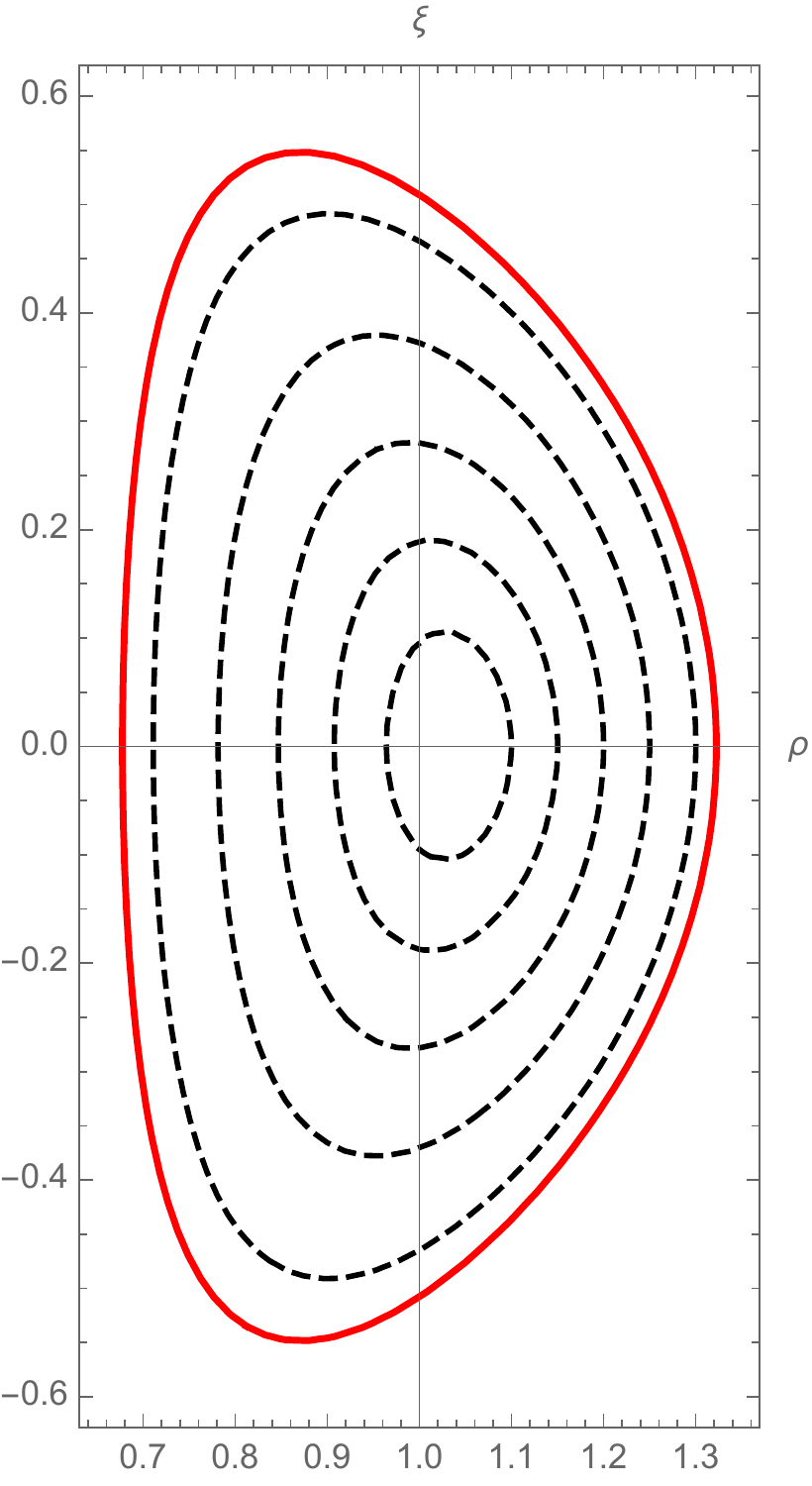}
\caption{A D-shaped equilibrium configuration constructed by means of our analytic solution with the following geometric characteristics: $\kappa_u=1.7$, $\delta=0.4$, $a=2\,m$ and $R_0=6.2\,m$. The solid red line represents the imposed boundary.\label{fig_2}}
\end{center}
\end{figure}
\begin{figure}
\begin{center}
\includegraphics[scale=0.55]{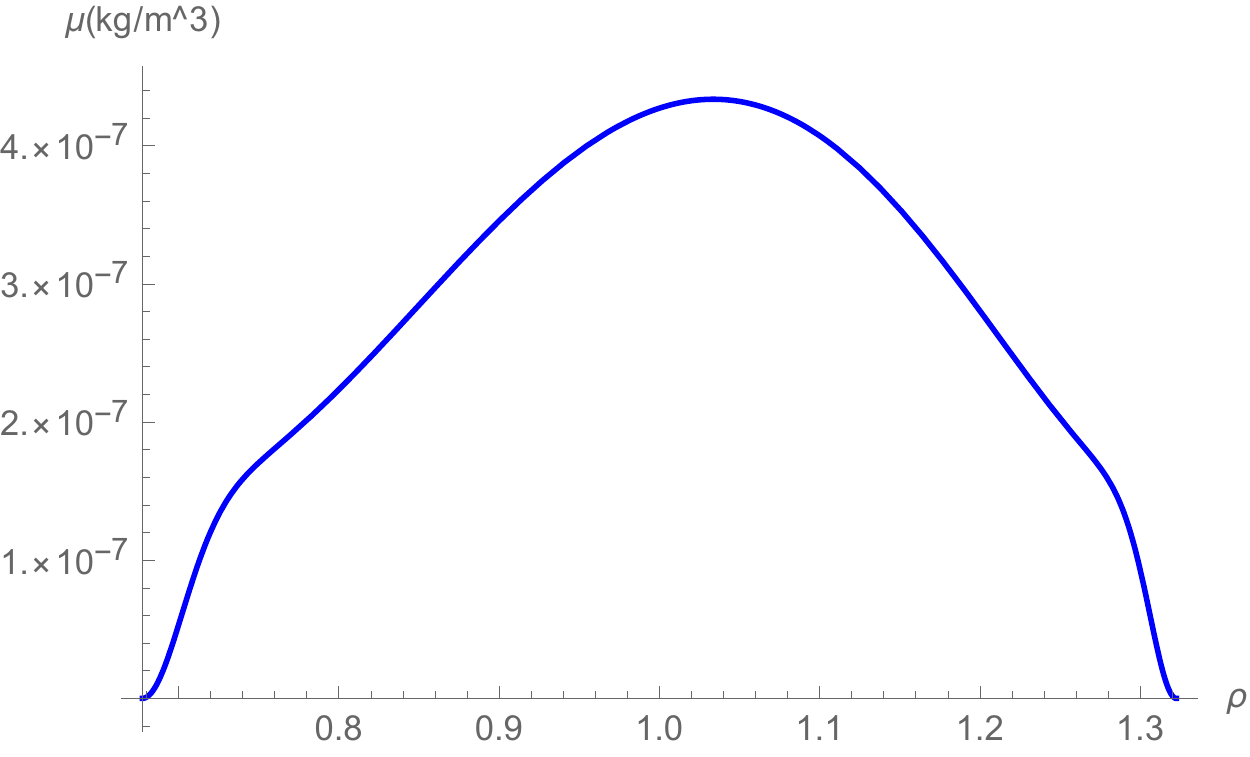}
\includegraphics[scale=0.55]{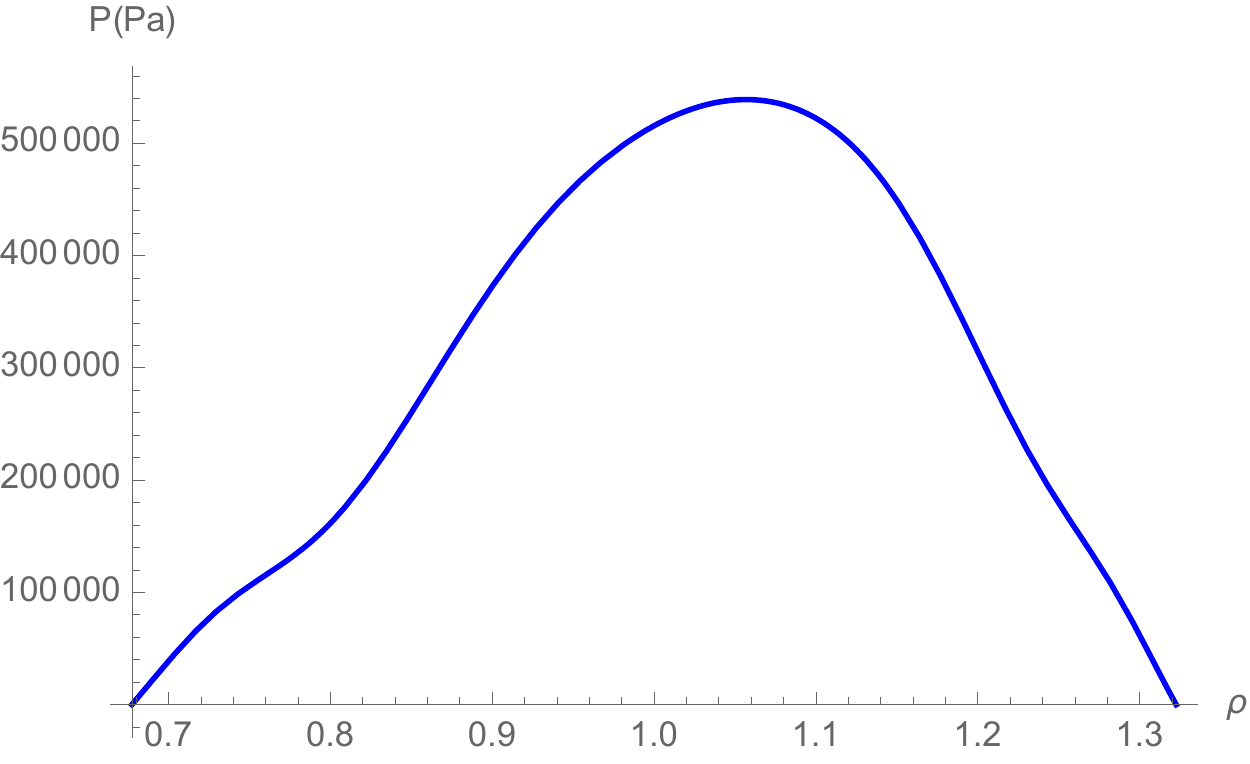}
\end{center}
\caption{An H-mode consistent density profile in connection with \eqref{den} and the pressure profile modified by the sheared flow. \label{fig_3}}
\end{figure}
\begin{figure}
\begin{center}
\includegraphics[scale=0.55]{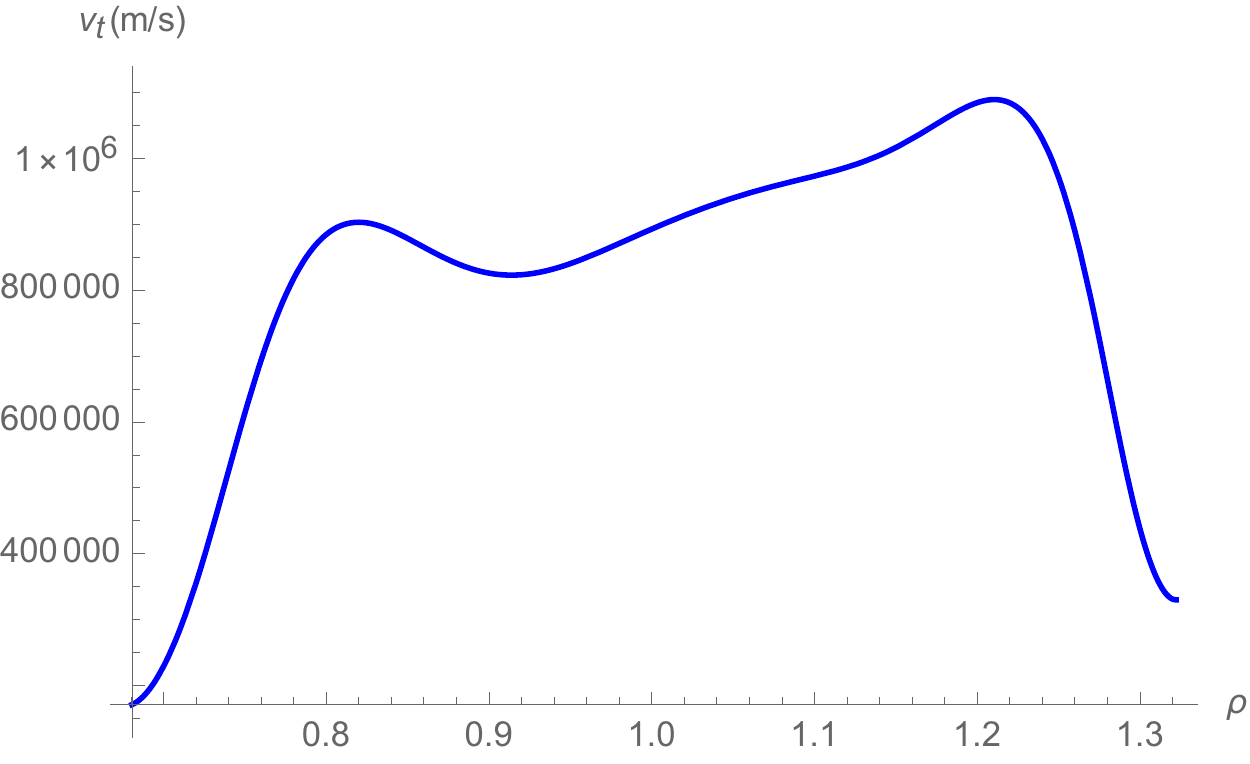}
\includegraphics[scale=0.55]{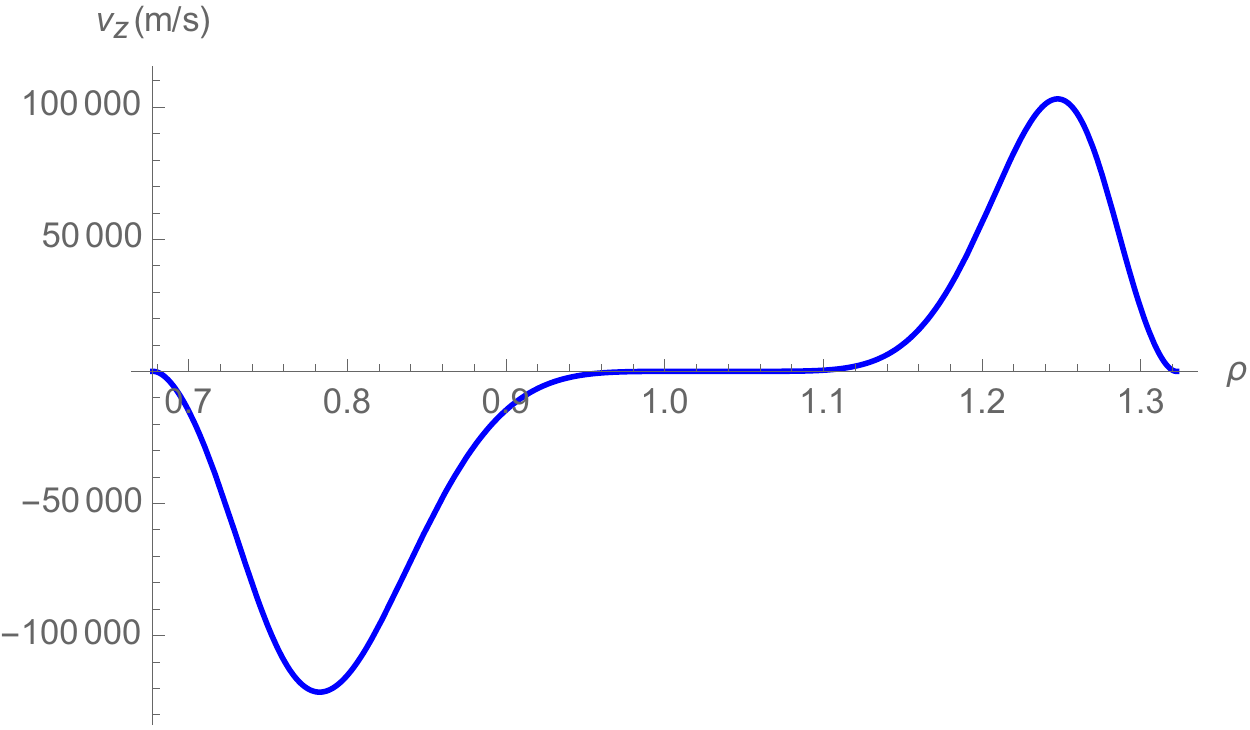}
\end{center}
\caption{The toroidal and the $z$ component of the flow velocity on $z=0$ in connection with \eqref{Mp}. \label{fig_4}}
\end{figure}

In case the solution contains more free parameters, we can determine the rest of them by imposing $u(\rho,\xi)=0$ on a set of boundary points lying in between the four characteristic points described above. To ensure that the Kruskal criterion $q>1$ is satisfied we employed the following formula for the safety factor on the magnetic axis ($\rho_a,\xi_a$):
\begin{equation}
q_a=\left\lbrack \frac{I}{\rho\sqrt{1-M_p^2}}\left(\frac{\partial^2 u}{\partial\rho^2}\frac{\partial^2u}{\partial\xi^2}\right)^{-1/2}\right\rbrack_{\rho=\rho_a,\xi=\xi_a}
\label{safa}
\end{equation}
 Then we computed  one of these free parameters upon requiring that the safety factor on the magnetic axis attains a value slightly above unity. By doing so, we were able to construct a tokamak, up-down  symmetric equilibrium configuration, with geometric characteristics compatible with those of ITER. The magnetic surfaces on a poloidal cross section are depicted in figure \ref{fig_2}.  To completely define the equilibrium state one should choose additionally the functional form of the free functions $\mu=\mu(u)$ and $M_p=M_p(u)$. For this particular example we adopted a choice relevant to the high confinement mode, where a mass density pedestal is formed and sheared flows are present. Namely, we chose
\begin{eqnarray}
\mu(u)&=&(\mu_0+\mu_1u^2)(1-e^{-u/\mu_2}) \label{den}\\
M_p^2&=&M_a^2\left(\frac{mu_a}{m+n}\right)^{-m}\left(\frac{nu_a}{m+n}\right)^{-n}\times\nn\\
&&\times (u_a-u)^m u^n\label{Mp}
\end{eqnarray} 
where $\mu_0$, $\mu_1$, $\mu_2$, $m$, $n$ are free parameters that are chosen appropriately in order to form the desired profiles.  Figures \ref{fig_3} and \ref{fig_4} depict some equilibrium quantities of interest featuring H-mode characteristics e.g. steep mass density gradients and high flow shear towards the boundary in connection with the formation of edge transport barriers. Note that if the flows were parallel to the magnetic field then the profile of the toroidal velocity should exhibit similar behavior with the poloidal velocity profile since they would both depend exclusively on the total Mach function $M$. Therefore, the presence of the $\rho^4$ term in \eqref{ggs}, associated with the non-parallel flow component, allows us to adjust the profiles of the toroidal and  poloidal velocities in different ways giving an additional degree of freedom and greater flexibility in constructing flowing equilibria.

In summary, adopting a generic linearizing ansatz for the free functions appearing in the GGSE \eqref{ggs},  describing axisymmetric states with incompressible flow of arbitrary direction, we solved analytically the resulting equation \eqref{ggs2} by an algorithm based on power series solutions.  Subsequently, we constructed a D-shaped tokamak stationary state with ITER pertinent geometric characteristics and smooth boundary. The solutions can be utilized  for testing the accuracy of equilibrium codes and furthermore they can potentially be exploited in constructing more realistic tokamak equilibria in connection with experimental measurements.

This work has been carried out within the framework
of the EUROfusion Consortium and has received funding
from the National Programme for the Controlled Thermonuclear
Fusion, Hellenic Republic. The views and
opinions expressed herein do not necessarily reflect those
of the European Commission. D.A.K. was financially
supported by the General Secretariat for Research and
Technology (GSRT) and the Hellenic Foundation for Research
and Innovation (HFRI).


\end{document}